\documentclass[aps,pre,amsmath,twocolumn,superscriptaddress]{revtex4-1}
\usepackage[figuresright]{rotating}
\usepackage{braket}
\usepackage{bm}
\usepackage{dsfont}
\usepackage{amsmath,amssymb}
\usepackage{color}
\usepackage{url}
\usepackage{algorithm}
\usepackage{algpseudocode}
\usepackage[titletoc]{appendix}
\usepackage{bbm}
\usepackage{bbold}

\DeclareMathOperator{\var}{var} 
\DeclareMathOperator{\tr}{tr}  
\newcommand{\im}{{i}}          
\newcommand{\id}{\mathbb{1}}   
\newcommand{\St}{\mathcal{S}}  

\begin{document}
\title{Computation of the asymptotic states of modulated open quantum systems \\
with a numerically exact realization of the quantum trajectory method}

\author{V.~Volokitin}
\affiliation{Mathematical Software and Supercomputing Technologies Department, Lobachevsky State University of Nizhny Novgorod, Russia}
\author{A.~Liniov}
\affiliation{Mathematical Software and Supercomputing Technologies Department, Lobachevsky State University of Nizhny Novgorod, Russia}
\author{I.~Meyerov}
\affiliation{Mathematical Software and Supercomputing Technologies Department, Lobachevsky State University of Nizhny Novgorod, Russia}
\author{M.~Hartmann}
\affiliation{Institut f\"ur Physik, Universit\"at Augsburg, Universit\"atsstra{\ss}e 1, D-86135 Augsburg, Germany}
\author{M.~Ivanchenko}
\affiliation{Department of Applied Mathematics, Lobachevsky State University of Nizhny Novgorod, Russia}
\author{P.~H\"{a}nggi}
\affiliation{Institut f\"ur Physik, Universit\"at Augsburg, Universit\"atsstra{\ss}e 1, D-86135 Augsburg, Germany}
\author{S.~Denisov}
\affiliation{Department of Applied Mathematics, Lobachevsky State University of Nizhny Novgorod, Russia}
\affiliation{Institut f\"ur Physik, Universit\"at Augsburg, Universit\"atsstra{\ss}e 1, D-86135 Augsburg, Germany}

\begin{abstract}
Quantum systems out of equilibrium are presently a subject of active research, 
both in theoretical and experimental domains. 
In this work we consider time-periodically modulated quantum systems which are 
in contact with a stationary environment. Within the framework of a quantum master 
equation, the asymptotic states of such systems are described by time-periodic density operators. 
Resolution of these operators constitutes a non-trivial computational task. 
Approaches based on spectral and iterative methods are restricted to systems 
with the dimension of the hosting Hilbert space dim$\mathcal{H} = N\lesssim 300$, while 
the direct long-time numerical integration of the master equation becomes increasingly problematic for $N \gtrsim 400$, especially when the coupling to the environment is weak. 
To go beyond this limit, we use the quantum trajectory method which unravels  master equation 
for the density operator into a set of stochastic processes for wave functions. The asymptotic density matrix is calculated by performing 
a statistical sampling over the ensemble of quantum trajectories, preceded by a long transient propagation.  We follow the ideology of event-driven programming
and  construct a new algorithmic realization of the method. The algorithm is computationally efficient, allowing for long 'leaps' forward in time, and is numerically
exact in the sense that, being given the list of uniformly distributed (on the unit interval) random numbers,
$\{\eta_1, \eta_2,...,\eta_n\}$, one could propagate a quantum trajectory (with $\eta_i$'s as norm thresholds) in a numerically exact way.
By using a scalable $N$-particle quantum model, we demonstrate that the algorithm allows us to resolve the asymptotic density operator 
of the model system with $N = 2000$ states on a 
regular-size computer cluster, thus reaching the scale on which numerical studies of modulated  Hamiltonian systems are currently performed.
\end{abstract}

\maketitle

\section{Introduction}

Most of \textit{in vivo} quantum systems are interacting with an environment. Although 
often weak, this interaction becomes relevant when studying the evolution of a system over long time scales. In particular,
the asymptotic state of such an \textit{open} system depends both on the unitary
action induced by the system Hamiltonian, and the action of the environment,
conventionally termed `dissipation'. A recent concept of
``engineering by dissipation'' \cite{kraus,barr,kienz,inf,inf1}; i.e., the creation of
designated pure and highly entangled states of many-body quantum systems by
using specially designed dissipative operators, has promoted the role of quantum dissipation to the
same level of importance as unitary evolution.

The use of time-periodic modulations constitutes another means to manipulate the
states of a quantum system. In the coherent limit, when the system is decoupled
from the environment, the modulations imply an explicit time-periodicity of the
system Hamiltonian, $H(t+T) = H(t)$. The dynamics of the system are determined
by the basis of time-periodic \textit{Floquet eigenstates}
\cite{shirley,sambe,grifoni,Eckardt2015}. The properties of the Floquet states depend on
various modulation parameters. Modulations being resonant with intrinsic
system frequencies may create a set of non-equilibrium eigenstates with
properties drastically different from those of time-independent Hamiltonians. Modulations enrich 
the physics occurring in fields such as quantum optics, optomechanics, solid state and ultra-cold atom physics
\cite{grifoni,kohler,Eckardt2015,bukov} and disclose a whole  spectrum of new phenomena
\cite{gong,top,majorana,gauge,Eckardt2017}.

What are the possible physical prospects of a synergy between
environment-induced decoherence and periodic modulations when both aspects  impact a
$N$-state quantum system? Of course, this question should be rephrased more precisely, depending on
the context of the problem. However, we are confident that a partial answer to this
question, even in his most general form,  will be appreciated by several communities working on
many-body localization (MBL) \cite{alt,abanin,lazarides,fish,les}, Floquet
topological insulators \cite{top}, and dissipative engineering
\cite{kraus,barr,kienz}.

There exist several approaches to address the evolution of open quantum
systems \cite{book}. A popular (especially  in the context of  quantum
optics \cite{carm}) approach  is based on the  quantum master equation with the generator
$\mathcal{L}$ of the Lindblad form \cite{lind,alicki} (we set $\hbar=1$):
\begin{eqnarray}
\nonumber
\dot{\varrho} = \mathcal{L}(\varrho) = -i[H(t),\varrho] + \sum_{k=1}^K \gamma_k(t)\cdot\mathcal{D}_k(\varrho),~~~~~~~~~~~~ \\~~~~~~~
\mathcal{D}_k(\varrho) = V_k\varrho V^\dagger_k
- \frac{1}{2}\{V^\dagger_kV_k,\varrho\}.~~~~~~~~~~~~~~~~~~~~~~~~~~~~~
\label{lind}
\end{eqnarray}
Here, $\varrho$ denotes the system density matrix, while the set of quantum
jump operators, $V_k$,
$k=1,...,K$, capture the action of the environment on the system. The jump operators act on the coherent
system dynamics via $K$
`channels' with time-dependent rates $\gamma_k(t)$. Finally, $[\cdot , \cdot]$ and $\{\cdot,\cdot\}$ denote the commutator and the
anti-commutator, respectively.

As an object of mathematical physics,
Eq.~(\ref{lind}) exhibits a specifically tailored structure and possesses a variety of important
properties \cite{alicki}. In the case of a time-independent, stationary
Hamiltonian $H(t) \equiv H$, the generator $\mathcal{L}$ induces a
continuous set of completely positive quantum maps
$\mathcal{P}_t = e^{\mathcal{L}t}$. Under some
conditions, the system evolves from an initial state
$\varrho^{\mathrm{init}}$
towards a unique and
time-independent asymptotic state $\varrho^{\mathrm{eq}}$,
$\lim_{t \rightarrow \infty}
\mathcal{P}_t\varrho^{\mathrm{init}} = \varrho^{\mathrm{eq}}$
\cite{alicki}. When time-periodic modulations are present,
Eq.~(\ref{lind}) preserves the complete positivity of the time evolution if
all those coupling rates are non-negative at any instance of time, $\gamma_k(t)
\geq 0,~~ \forall t$ \cite{alicki}. Under certain, experimentally
relevant assumptions an approximation in terms of a ``time-dependent
Hamiltonian and a time-independent dissipation'' provides a suitable approximation
\cite{alicki}.

Here, we address the particular case of quench-like, periodic
modulations of period $T$ with the time-periodic dependence of the Hamiltonian
$H(t)$, $t \in [0,T]$, consisting of a switch between several constant
Hamiltonians. A common choice is a  setup composed
of two Hamiltonians,
\begin{equation}
H(t) =
\left\{
  \begin{array}{cc}
   H_1, & \text{for}~~ 0 \leqslant     t \bmod T < \tau \\
   H_2, & \text{for}~~  \tau \leqslant t \bmod T < T
  \end{array}
\right. ,
\label{pc}
\end{equation}
with $\tau \in [0,T]$. This
minimal form has recently been used to investigate the connection between
integrability and thermalization \cite{lazarides,lazaridesT,rigol} or, alike,
for disorder-induced localization \cite{abanin}  in \textit{coherent}
periodically modulated many-body systems.

From a mathematical point of view, Eqs.~(\ref{lind},\ref{pc}) define a linear
operator equation with a time-periodic generator $\mathcal{L}(t)$. Therefore,
Floquet theory applies and asymptotic solutions of the equation are
time-periodic with temporal period $T$ \cite{grifoni,yakub}. $\mathcal{L}(t)$
is a dissipative operator and, in the absence of relevant symmetries, the system evolution in the asymptotic limit $t \rightarrow
\infty$ is determined by a unique `quantum attractor', i.e., by an asymptotic,
time-periodic density operator obeying $\varrho^\mathrm{att}(\tau + nT) =
\varrho^\mathrm{att}(\tau)$, $\tau \in [0,T]$ and $n \in \mathbb{Z}^+$. The
main objective here consists in explicit numerical evaluation of the matrix form of this operator.

To use spectral methods (complete/partial diagonalization
and different kinds of iterative algorithms \cite{nation1}) to calculate
$\varrho^\mathrm{att}$ as an eigenelement of the corresponding Floquet
map $\mathcal{P}(T) =
e^{\mathcal{L}_2(T-\tau)}e^{\mathcal{L}_1\tau}$
would imply that one has to deal with $N^2$  computationally expensive operations.
In the case of periodically modulated systems it restricts the use of spectral methods to $N \lesssim 300$ \cite{foot1}.

 A direct
propagation of Eq.~(\ref{lind}) for a time span long enough for $\varrho(t)$ to
approach the quantum attractor is not feasible for $N \gtrsim 400$ for at least two reasons:
Direct propagation requires to numerically propagate $N^2 \gtrsim 1.6 \cdot 10^5$
complex differential equations with time-dependent coefficients, so that
the accuracy might become problematic for large evolution times. Although the accuracy may be
improved by implementing high(er)-order integration schemes
\cite{qutip} or Faber and Newton polynomial integrators \cite{kosloff}, this
approach is hardly parallelizable so one could not benefit by propagating equations on a cluster \cite{foot2}.

Systems containing $N=400$ states may still be too small, for example, to explore MBL
effects in open periodically-modulated systems. Is it possible to exceed
this limit? If so, to what extent is this feasible? We attempt to answer these two questions by first 
unraveling  of the quantum master
equation (\ref{lind}) into a set of stochastic realizations, by resorting to the celebrated method of ``quantum
trajectories'' \cite{zoller,dali,plenio,daley}. This method allows one to
transform the problem of the numerical solution of Eqs.~(\ref{lind}-\ref{pc})
into a task of statistical sampling over quantum trajectories which form
vectors of the size $N$. The price to be paid  for the reduction from $N^2$ to $N$
is that we now have to sample over many realizations. This problem is very well
suited for parallelization and we thus can  benefit from the
use of a computer cluster. If the number of realizations $M_{\mathrm{r}}$
becomes large, the sampling of the density operator $\varrho(t)$ with the initial
condition $\varrho^{\mathrm{init}}= |\psi^{\mathrm{init}}\rangle\langle
\psi^{\mathrm{init}} |$ converges to the solution of Eq.~(\ref{lind})
\cite{zoller,dali} provided that the propagation of the trajectories was perform \textit{in the exact way} (we discuss the precise meaning of this in Section III). 

We address the generic system Eqs.(\ref{lind}-\ref{pc}), with no conditions imposed
on the operators $H(t)$ and $V_k$ (for example, they need not be local
\cite{cirac1,vidal} and with no {\it a priori} knowledge of the attractor state.
The are two important issues. First is the time  $t_{\mathrm{p}}$ after which the trajectories are
sampled. To guarantee that the asymptotic regime is reached, this time has to
exceed the longest relaxation timescale of the system. Practically, this means
that the sampling over trajectories started at time $t_{\mathrm{p}} =ST$, with integer $S \gg 1$,
does converge to a density operator, which is close to the asymptotic
$\varrho^\mathrm{att}(\tau = 0)$. Second, in order to minimize numerical errors due to long
propagation, we devise an integration scheme based on a set of exponential
propagators. For quench-like periodic modulations this implies a finite number
of propagators which can be pre-calculated and stored locally on each cluster
node, as we discuss in the next section. 

For a scalable model, a periodically rocked and
dissipative  dimer with $N-1$ interacting bosons, we find that the
statistical variance of the sampling does not grow infinitely with
$t_{\mathrm{p}}$ but rather saturates to a limit-cycle evolution. Therefore,
the number of trajectories $M_{\mathrm{r}}(\epsilon)$ needed to estimate
elements of $\varrho^\mathrm{att}$ with accuracy $\epsilon$ (defined with some matrix norm), remains finite.
Assuming that the propagation can be performed for an arbitrary large time
$t_{\mathrm{p}}$ with required accuracy, we are left with the only  problem to
sample over a sufficiently large number of trajectories.

In addition, in the asymptotic limit, the sampling of $\varrho^\mathrm{att}(\tau = 0)$ can be
performed over individual trajectories stroboscopically, after each period $T$.
This increases the efficiency of sampling via the use of the same trajectory
without having to initiate yet a new trajectory and then propagating it up to
time $t_{\mathrm{p}}$. Our results confirm that by implementing this approach
on a cluster, it is possible to resolve attractors
of periodically modulated open systems with several thousand quantum states,
thus increasing $N$ by one order of magnitude.

The present work is organized as follows: In Section~II we outline
the method of quantum trajectories and describe the algorithmic realization of the method. 
Statistical aspects of sampling are briefly discussed in
Section~III. In Section~IV we introduce a scalable model system which 
serves as a testbed for the algorithm. Section~V is devoted to the implementation of the algorithm on a cluster together with an analysis of its performance and
scalability. Section~VI reports numerical results obtained for the
test case. The findings of the study are summarized together with an outline of further
perspectives in the final Section~VII.

\section{Quantum trajectory as an event-driven process}

To sample the solution of Eqs.~(\ref{lind} - \ref{pc}) up to some time $t_{\mathrm{p}}$ using
quantum trajectories (also known under the labels of quantum jump method
\cite{plenio} or the Monte Carlo wave function method \cite{dali}) we
first have to calculate the effective non-Hermitian Hamiltonian
\begin{eqnarray}
\tilde{H}(t) = H(t) -\frac{i}{2} \sum_{k=1}^K V^\dagger_k V_k ,
\label{nonH}
\end{eqnarray}
and then proceed along the following path of instructions \cite{zoller}:
\begin{enumerate}

\item initiate the trajectory in a pure state $\ket{\psi^{\mathrm{init}}}$;

\item draw a random number $\eta$ which is uniformly distributed on the unit interval;

\item propagate the quantum state $\ket{\psi(t)}$ in time using the effective Hamiltonian $\tilde{H}(t)$;

\item the squared norm $\lVert  \ket{\psi(t)} \rVert^2$ decays monotonically. 
When the equality $\eta = \lVert \ket{\psi(t)} \rVert^2$ is reached, stop the propagation and normalize the state vector,  $\ket{\psi(t)} \rightarrow \ket{\psi(t)}/\lVert \ket{\psi(t)} \rVert$;

\item perform a quantum jump: select the jump operator $D_k$ with probability
     $p_k = \gamma_k \lVert D_k\ket{\psi(t)} \rVert^2/\sum_{k=1}^K \gamma_k \lVert D_k\ket{\psi(t)} \rVert^2$ and apply the transformation
     $\ket{\psi(t)} \rightarrow D_k\ket{\psi(t)}/\lVert D_k\ket{\psi(t)} \rVert$;

    \item repeat steps 2 -- 5 until the desired time $t_{\mathrm{p}}$ is reached.
\end{enumerate}

The density matrix can then be sampled from a set of $M_r$ realizations as
$\varrho(t_\mathrm{p};M_{\mathrm{r}}) = \frac{1}{M_{r}}
\sum_{j=1}^{M_{\mathrm{r}}} \ket{\psi_j(t_\mathrm{p}}\bra{\psi_j(t_\mathrm{p})}$.
Formally, in the limit $M_{\mathrm{r}} \rightarrow \infty$,
the result $\varrho(t_\mathrm{p};M_{\mathrm{r}})$ converges towards the solution of Eq.~(\ref{lind}) at time $t_\mathrm{p}$ for the given
initial density matrix $\varrho^{\mathrm{init}} = \ket{\psi^\mathrm{init}}\bra{\psi^\mathrm{init}}$
\cite{book,zoller}. The density matrix can also
be sampled at any other instance of time $t \in [0,t_\mathrm{p}]$. This does
not affect the propagation of the trajectory and only demands normalization
of the state vector $\ket{\psi(t)}$ before updating
$\varrho(t;M_{\mathrm{r}}) \rightarrow \varrho(t;M_{\mathrm{r}}+1)$. More
specifically, an element of the density matrix, $\varrho_{ls}(t)$, should be
sampled as 
\begin{equation}
\label{eq:4a}
\varrho_{ls}(t;M_{\mathrm{r}}) = \frac{1}{M_{r}} \sum_{j=1}^{M_{\mathrm{r}}} c_{j,l}(t)\cdot c_{j,s}^{\dagger}(t),
\end{equation}
where $c_{j,l}(t)$ is the $l$-th coefficient of the expansion (in the same basis $\{\ket{\psi_j(t)}\}, k=1,\ldots,N$ used to express the density matrix) of the normalized wave-function,
$\ket{\psi_j(t)} = \sum_{l=1}^{N} c_{j,l}(t) \ket{k}$.

The recipe contains two key steps: (i) propagation (step 3) and (ii) determination of the time of the next
jump (step 4). The waiting time, i.e., the time between two consecutive jumps,
cannot be obtained without actual propagation of the trajectory (except in a
few cases \cite{book,zoller}). This time must be obtained along with the
numerical integration by using the non-Hermitian Hamiltonian $\tilde{H}(t)$.
One has to propagate a trajectory, monitor the decaying squared
norm of the wave vector and determine the instant of time when the squared norm
equals the randomly chosen value $\eta$. In most of the existing studies, this was realized with a
step-by-step Euler method. This approach, although having a
physical interpretation \cite{zoller}, is not suitable for our purpose because
it corresponds to the expansion of Eq.~(\ref{lind}) to the first order in a time
step $\delta t$; consequently, a reasonable accuracy of the sampling can be
achieved with extremely small values of $\delta t$ only \cite{foot3}.

Several improvements based on higher-order (with respect to $\delta t$) unraveling schemes
\cite{high1,high2} have been put forward. The accuracy of the sampling -- for the same number of
realizations $M_{\mathrm{r}}$ and time step $\delta t$ -- can be improved
substantially by increasing the order of the integration scheme \cite{high1}. In
QuTiP, an open-source toolbox in Python to simulate dynamics of open
quantum systems \cite{qutip}, Adams method (up to $12$-th order) and backward
differentiation formula (up to fifth order) with adaptive time step are
implemented. In this respect, this is presently the most advanced implementation, to the best of our knowledge. In addition, QuTiP supports time-dependent Hamiltonians and allows for
multi-processor parallelization.  
The original publication \cite{qutip} 
addressed scalability and performance of the QuTiP 
package and demonstrated that a stationary model with $N=8000$ states can be propagated. 
However, the results remained restricted to averaging over a few quantum trajectories and relatively short propagation 
time $t_{\mathrm{p}}$. Also, the issues of accuracy and convergence to an asymptotic state with the number of sampled trajectories 
were not discussed. 

In contrast, aside of reaching large $N$, we are concerned about the following two issues. 
First, there is the accuracy of propagation. As $t_{\mathrm{p}}$ has to be extremely large in order to 
be able to sample a state close to the attractor state $\varrho^\mathrm{att}$ (note that up to now the method of
quantum trajectories was used mainly to analyze short-time relaxation and
transient regimes in terms of some observables; e.g.\ see in Refs.~\cite{les}), the accumulating error
due to the discrete approximation of the continuous evolution with
the effective Hamiltonian $\tilde{H}$ can emerge sizable. These errors may cause
serious problems, for example, when dealing with the delicate issue of MBL
phenomena. Second, in the limit of weak dissipation, when the coupling rates $\gamma_k$ are
small, jumps occur rarely. For most of the time the
evolution of the trajectory is deterministic and propagation using a small
$\delta t$ will not be efficient. Increasing the time step implies
a decrease of the accuracy of determination of the time of the
jump. This constitutes yet another factor which can blur the quality of the
sampling scheme. On the other side, we want to maximize the speed (in terms of computational time) of the propagation.
If these two problems are successfully overcome, the only
remaining problem left is to obtain a sufficiently large number of
realizations. Here, we handle both issues with an approach presenting
an alternative to the schemes which rely on increasing the
order of integration.

A quantum trajectory serves a typical example of a so-called ``event-driven
processes'' used in control theory \cite{control} (where they are also known as  ``Lebesgue sampling processes'') and likewise also in
computational neuroscience \cite{neiro}. The question how to integrate such
processes \textit{numerically exact} has been discussed in those  research
areas already since the late $1990$s. A possible option consists in the combination of
an exponential propagation together with time-stepping techniques. We next mainly follow the
idea put forward with Ref.~\cite{neiro}.

To start, let us first  consider a time-independent Hamiltonian $H(t) \equiv H$. The propagation
over any time interval $\delta t$ with the corresponding effective Hamiltonian
$\tilde{H}$ can be done by the propagating operator (propagator)
$P_{\delta t} = e^{-i\tilde{H}\delta t}$. Exponentiation of $\tilde{H}$ can be
performed numerically with controllable accuracy by implementing the scaling
and squaring method \cite{exp}. To determine the time of the next jump, we use
a time stepping technique \cite{neiro}. We choose the convenient and efficient
bisection method \cite{knuth}, cf. Section V for more details. The
accuracy of the bisection method is controlled by the maximal order of
bisections $S$ which we call `maximal depth'. The time of the jump is thus
resolved with a precision $2^{-S}\delta t$. A practical realization of this
method demands a set of $S$ propagators, that is, $P_{\delta t_s} =
e^{-i\tilde{H}\delta t_s}$, $\delta t_s = 2^{-s}\delta t_0$, $s=0,...,S$, that
are complex $N \times N$ matrices. These propagators have to be pre-calculated
and then stored. Generalization of this approach to the case of quench-like
temporal modulations is straightforward. In the bi-Hamiltonian case, i.e.,
Eq.~(\ref{pc}), we have to double the number of the stored propagators and then
switch between the two sets every half of the temporal period $T$.

Our key objective here is to estimate the maximal
system size $N$ for Eqs. (\ref{lind} - \ref{pc}), whose asymptotic density matrix
can be resolved with quantum trajectories implemented on a (super)computer cluster and validate the accuracy of the sampling algorithm. 

\section{Statistical error(s) of sampling} 

We next  discuss the problem of statistical errors. We assume that the integration of quantum trajectories is performed
in a numerically exact way, i.e., when the list of consequent norm thresholds, $\{\eta_1, \eta_2,...,\eta_n\}$,
and the initial state, $\varrho^{\mathrm{init}}= |\psi^{\mathrm{init}}\rangle\langle
\psi^{\mathrm{init}} |$, are given, the corresponding trajectory can be calculated  numerically exact.

Consider the sampling of a
variable $X(t)$ over an ensemble of realizations $\big\{ X_j(t) \big\}$, $j =
1,...,M_{\mathrm{r}}$, with the aim to estimate its mean $\tilde{X}(t)$.
Examples would be the expectation value of an operator
\cite{zoller,dali,plenio,daley} or an element of the density matrix (as in our
case). In addition to the mean (average) of the variable,
$\tilde{X}(t;M_{\mathrm{r}}) = \frac{1}{M_{\mathrm{r}}}
\sum_{j=1}^{M_{\mathrm{r}}} X_j(t)$, we can also calculate its variance
\cite{book,daley},
\begin{eqnarray}
\var\big[X(t);M_{\mathrm{r}}\big] = \frac{1}{M_{\mathrm{r}}} \sum_{j=1}^{M_{\mathrm{r}}} \Big(X_j(t) - \tilde{X}(t;M_{\mathrm{r}})\Big)^2,
\label{variance}
\end{eqnarray}
which here for systems possessing a finite Hilbert space dimension $N$ is assumed to
converge to a generally time-dependent value $\var\big[X(t) \big]$ in the
limit $M_{\mathrm{r}}\rightarrow \infty$. Different trajectories are
statistical independent. Therefore, the central limit theorem applies and, for
large $M_{\mathrm{r}}$, the probability density function (pdf) of the mean
$\bar{X}(t;M_{\mathrm{r}})$ can be approximated by a Gaussian pdf centered at
$\tilde{X}(t)$ with the standard deviation $\sigma(t;M_{\mathrm{r}}) =
\sqrt{\var(X;M_{\mathrm{r}})/M_{\mathrm{r}}} \stackrel{M_{\mathrm{r}} \gg
1}{\propto} M_{\mathrm{r}}^{-\frac{1}{2}}$.

In the framework of local and global quantities \cite{dali}, elements of the
density matrix correspond to the former. That means that in order to resolve 
their values we need $M_{\mathrm{r}} \gg N$ realizations.
In addition, they are small for large
$N$, $\varrho_{kl} \sim \mathcal{O}(N^{-1})$, and the standard criterium of a
trustful sampling, $ \sigma(M_{\mathrm{r}})/\varrho_{kl} \ll 1$, implies that
$M_{\mathrm{r}} \gg N^2$. Such a massive sampling is unfeasible if $N \gtrsim
10^3$, even when used on a supercomputer. However, this constitutes a sufficient
condition which greatly overestimates (hopefully) the number of realizations
needed for a reasonable resolution of the density matrix, as we scrutinize for our test case below.
This presents yet another aspect of the sampling with quantum trajectories we aim to gain
more specific insight.

Another issue we like to mention is the time evolution of the variance
$\var[\varrho_{kl}(t)]$. Evidently, it cannot grow to infinity simply because
the absolute values of the coefficients $c_s(t)$
do not exceed one. Therefore, there is an upper limit $\var[\varrho_{kl}(t)] \simeq 1$.
On the other hand, for completely random and uniformly distributed values of
$c_s(t)$ we find $\var[\varrho_{kl}(t)] \propto N^{-1}$. By using a scalable
model we show that (i) the variances saturate in course of propagation to
time-periodic values, $\var[\varrho_{kl}(t+T)] = \var[\varrho_{kl}(t)]$, which in
addition (ii) allow for an accurate estimation of the density matrix elements
with less than $N^2$ realizations.

\section{A model} 

As a testbed for the algorithm we use an open  physical system made up of $N-1$ indistinguishable interacting
bosons which hop between a periodically rocked dimer. The system
Hamiltonian rusingeads explicitly:
\begin{align}
H(t)=&-J \left( b_1^{\dagger} b_2 + b_2^{\dagger} b_1 \right) + \frac{U}{2(N-1)} \sum_{g=1,2} n_g\left(n_g-1\right)\nonumber \\
&+\varepsilon(t)\left(n_2 - n_1\right) \;. \label{eq:Hamiltonian}
\end{align}
Here, $J$ denotes the tunneling amplitude, $U$ is the interaction strength,
and $\varepsilon(t)$ presents a periodically varying modulation of
the local potential in time. In particular, we choose
$\varepsilon(t)=\varepsilon(t+T)=\mu_0+\mu_1 Q(t)$, where $\mu_0$ and $\mu_1$
denote a static and a dynamically varying, respectively,  energy offset between the two sites.
$Q(t)$ itself is a periodically varying, unbiased two-valued quench-function within
one full period $T$; more specifically, $Q(\tau) = \frac{1}{2}$ within $0 < \tau \le T/2$
and $ Q(\tau) = -\frac{1}{2}$ for the second half period $ T/2 < \tau \leq T$.
The boson operators $b_g$ and $b_g^{\dagger}$ are the annihilation and creation
operators on site $g \in \{1,2\}$, while $n_g=b_g^{\dagger}b_g$ is the particle
number operator. The system Hilbert space has dimension $N$ and can be spanned
with the $N$ Fock basis vectors, labeled by the number of boson on the first
site $n$, $\{|n+1\rangle\}$, $n=0,...,N-1$. Thus, the model size is controlled
by the total number of bosons. The Hamiltonian (\ref{eq:Hamiltonian}) has been
used for theoretical studies before in Refs. \cite{Vardi, Witthaut, PolettiKollath2012} and, as well, has
been implemented in recent experiments \cite{oberthaler, ober1}.

For the single jump operator we use \cite{DiehlZoller2008},
\begin{equation}
V=(b_1^{\dagger} + b_2^{\dagger})(b_1-b_2), \label{eq:jump}
\end{equation}
which attempts to `synchronize' the dynamics on the sites by constantly
recycling anti-symmetric out-phase modes into symmetric in-phase ones. The
dissipative coupling constant $\gamma = \gamma_0/(N-1)$ is taken  to be
time-independent. Since the jump operator is non-Hermitian, the propagators
$\mathcal{P}_t$ are not unital and the attractor does not assume the maximally
mixed state, $\varrho^\mathrm{att} \neq \id/N$.

The Hamiltonian (\ref{eq:Hamiltonian}) is non-integrable when $U \neq
0$; therefore, an analytical solution of the corresponding Lindblad equation
is not available. However, in the limit $N \rightarrow \infty$ the dynamics can be
approximated by mean-field equations for the expectation values of the three pseudo-spin
operators $\St_x=\frac{1}{2(N-1)}\left(b^{\dagger}_1 b_2 + b^{\dagger}_2
b_1\right)$, $\St_y= -\frac{\im}{2(N-1)}\left(b^{\dagger}_1 b_2 - b^{\dagger}_2
b_1\right)$, $\St_z=\frac{1}{2(N-1)}\left(n_1 - n_2\right)$. For a large number
of atoms, the commutator $\left[\St_x,\St_y\right]= [\im\St_z/(N-1)]
{\overset{N\to\infty}{=}} 0$ and similarly for other cyclic permutations.
Replacing operators with their expectation values, $\braket{\St_k} =\tr
[\varrho \St_k]$, and denoting $\braket{\St_k}$ by $S_k$, we find the
semi-classical equations of motion \cite{map}
\begin{align}
\frac{\mathrm{d} S_x}{\mathrm{d}t} &= 2\varepsilon(t)S_y - 2U S_zS_y + 8\gamma_0 \left(S_y^2+S_z^2\right), \nonumber\\
\frac{\mathrm{d} S_y}{\mathrm{d}t} &= -2\varepsilon(t)S_x + 2U S_xS_z +2JS_z - 8\gamma_0 S_xS_y, \nonumber\\
\frac{\mathrm{d} S_z}{\mathrm{d}t} &= -2JS_y - 8\gamma_0 S_xS_z.
\label{eq:mean}
\end{align}
As $S^2=S_x^2+S_y^2+S_z^2 = 1/4$ is a constant of motion, we
can reduce the mean-field evolution to the surface of a Bloch sphere,
$\left(S_x,S_y,S_z\right) = \frac{1}{2} [\cos(\varphi)\sin(\vartheta),\sin(\varphi)\sin(\vartheta), \cos(\vartheta) ]$,
yielding the equations of motion
\begin{align}
\dot{\varphi} &= 2J\frac{\cos(\vartheta)}{\sin(\vartheta)}\cos(\varphi) - 2\varepsilon(t) +U \cos(\vartheta) - 4\gamma_0 \frac{\sin(\varphi)}{\sin(\vartheta)}, \nonumber \\
\dot{\vartheta} &= 2J\sin(\varphi) + 4\gamma_0 \cos(\varphi)\cos(\vartheta) \label{eq:thetadot}\,.
\end{align}
The density matrix $\varrho$ of the system with $(N-1)$ bosons can be visualized
on the same Bloch sphere by plotting the Husimi distribution
$p(\vartheta,\varphi)$, obtained by projecting $\varrho$ on the set of the
generalized SU(2) coherent states, $\ket{\theta,\varphi} =
\sum_{j=0}^{N-1} \sqrt{{{N-1}\choose{j}}} \big [\cos(\theta/2) \big ]^j \big[e^{i\phi}\sin(\theta/2) \big ]^{N-1-j} \ket{j}$ \cite{coherent1,coherent2}.
The visual comparison of the Husimi distribution with the mean-field solution,
Eq.~(\ref{eq:thetadot}), will serve as a  test of the meaningfulness
of the sampled density matrix $\varrho(t_\mathrm{p};M_{\mathrm{r}})$.

\section{Implementation on a cluster and performance } 

Next we describe a high-performance implementation of the algorithm on a
supercomputer and analyze the scalability of its implementation by using the
model system (\ref{eq:Hamiltonian}, \ref{eq:jump}).
Numerical experiments were performed on the ``Lobachevsky'' supercomputer
\cite{lobachevsky} at the Lobachevsky State University of Nizhny Novgorod. We
employed up to $32$ computing nodes, with the following configuration per node:
$2 \times$ Intel Xeon E$5-2660$ CPU ($8$ cores, $2.2$ GHz), $64$ GB RAM, OS
CentOS $6.4$. We use Intel Math Kernel Library (MKL), Intel $\mathrm{C/C++}$
Compiler, and Intel MPI from Intel Parallel Studio XE \cite{IPSXE}.

\begin{table}[h]
\caption{Scaling efficiency on shared memory.
}
\label{table1}
\vspace{6pt}
\small
\centering
\begin{tabular}{c c c c c c c c c}
\hline
Number of threads & Time of computations,  & Efficiency, \\
                  & in seconds             & percent \\
\hline
\ 1  \ & 2170 & 100 \\
\ 2  \ & 1114 & 97 \\
\ 4  \ & 557  & 97 \\
\ 8  \ & 292  & 93 \\
\ 16 \ & 156  & 87 \\
\hline
\end{tabular}
\end{table}

Using Eq.~(\ref{nonH}), we start with two effective non-Hermitian
Hamiltonians, $\tilde{H}_1$ and $\tilde{H}_2$, describing the quench-like
modulations, Eq.~(\ref{pc}), as represented by a pair of complex
double-precision $N \times N$ matrices. An initial pure state
$|\psi^{\mathrm{init}}\rangle$ is represented by a complex-valued
double-precision vector. The propagation operator yields a wave function for a
single sample. We follow the straightforward approach to parallelization with
an independent random sampling. Namely, the computational load is distributed
among supercomputer nodes by the standard Message Passing Interface (MPI). On
each node we employ the OpenMP threads to parallelize sampling. Computationally
intensive operations are implemented by calling BLAS functions from Intel MKL
in sequential mode.

The code consists of three main steps. First, the program initializes MPI,
allocates memory, reads parameters and the matrices of the pre-calculated
exponential propagators from configuration files. The propagators are calculated independently
on each cluster node.
On the second step all OpenMP threads in all MPI processes independently
propagate several quantum trajectories starting from the initial state $|\psi^{\mathrm{init}}\rangle$ \cite{foot4}.

\begin{algorithm}[H]
\caption{: propagation of a quantum trajectory with exponential operators and bisection method}
\label{alg:AllBodies}
\begin{algorithmic}[1]
\State set $\delta t = \delta t_0$ $\&$ $s=0$
\While{$\lVert |\psi(t)\rangle \rVert^2 > \eta$}
\State calculate $|\tilde{\psi}(t)\rangle = P_{\delta t} |\psi(t)\rangle$
\If{$\lVert |\psi(t)\rangle \rVert^2 < \eta$ $\&$ $s < S$}
\State $s=s+1$
\Else
\State~~~$|\psi(t)\rangle = |\tilde{\psi}(t)\rangle$
\State~~~$t=t+\delta t$
\While{ $s > 0$ $\&$ $\delta t = n\cdot \delta t_{s-1}$, $n \in \mathbb{Z}^{+}$}
\State $s=s-1$; $\delta t = \delta t_s$\EndWhile
\EndIf
\EndWhile


\end{algorithmic}
\end{algorithm}

\begin{figure}[t]
\includegraphics[width=0.48\textwidth]{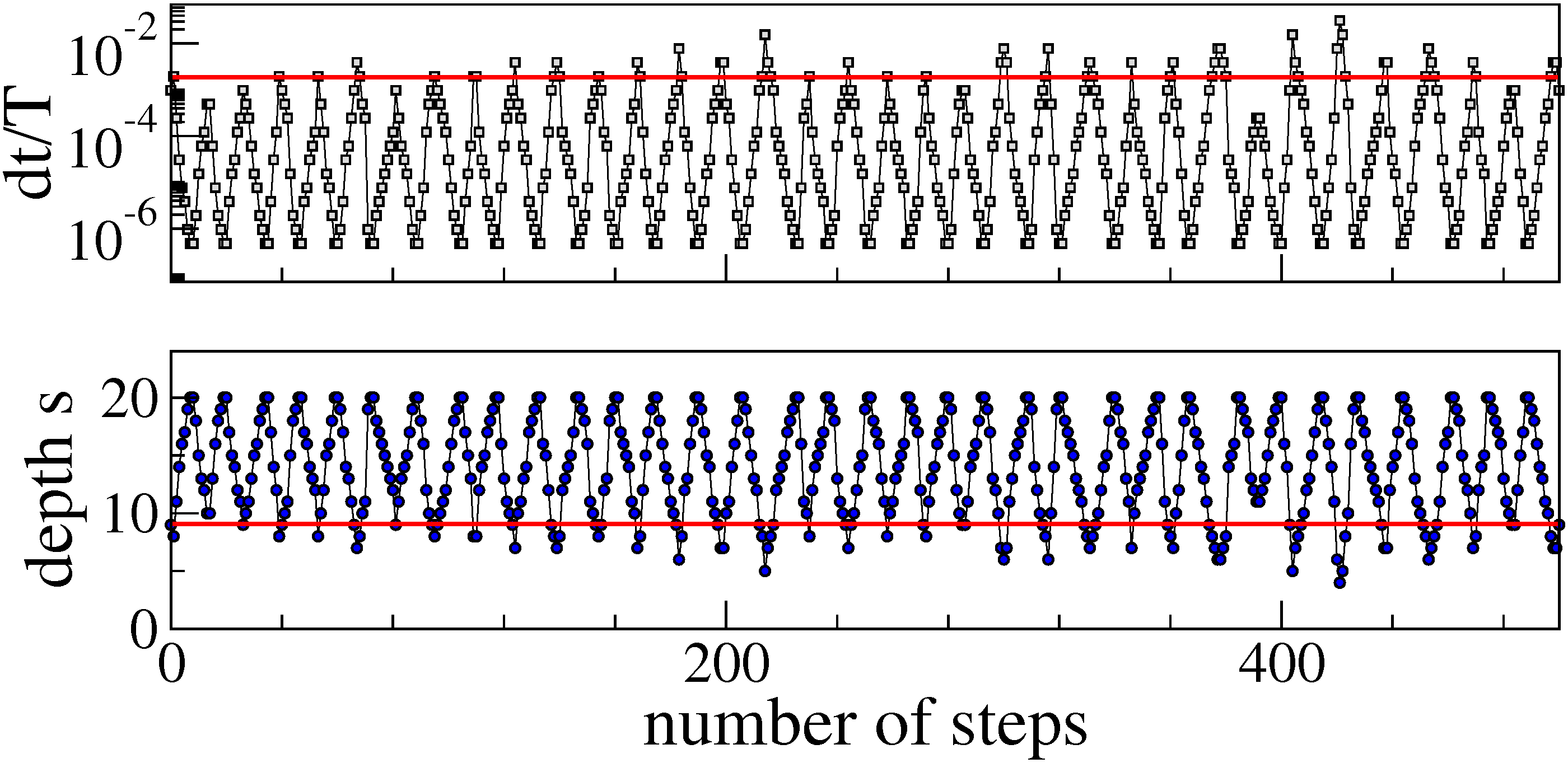}
\caption{(color online) Float-like performance of Algorithm \ref{alg:AllBodies} during the
propagation of a quantum trajectory. Every local maximum in the dependence
$dt$ vs number of steps (minimum in the depth $s$ dependence) indicates an
occurrence of a jump after which the algorithm performs a chain of bisections
to reach the maximal depth $S=20$. After every step during which no jump
occurred, the algorithm doubles the step size. The average time between two
consecutive jumps (red line) is the average height of the local maxim minus 1.
The two sequences were monitored during the sampling of the asymptotic state.}
\label{Fig:Husimi}
\end{figure}

The propagation is realized by using the
step-decimation technique. This pseudo-code
is presented in Algorithm 1. The maximal depth $S$, the time steps $\delta t_s = 2^{-s}dt$, and the
exponential propagators $P_{\delta t_s}$, $s=0,...,S$ are pre-loaded. The
program is initiated with $s=0$, but later on $s$ is taken from the previous
propagation loop step. This step is fully parallel; it contains a matrix-vector multiplication that is the 
most computationally intensive part of the algorithm. This operation is performed with the $zgemv$ MKL
subroutine. During the third step all samples on each node are accumulated into
the density matrix. Next, these matrices are collected in the rank $0$ MPI
process. Finally, one evaluates the resulting density matrix. This matrix is
written to the output file, the dynamic memory is deallocated and the MPI is
finalized.

\begin{figure*}[t!!!]
\begin{tabular}{cc}
\includegraphics[width=0.9\textwidth]{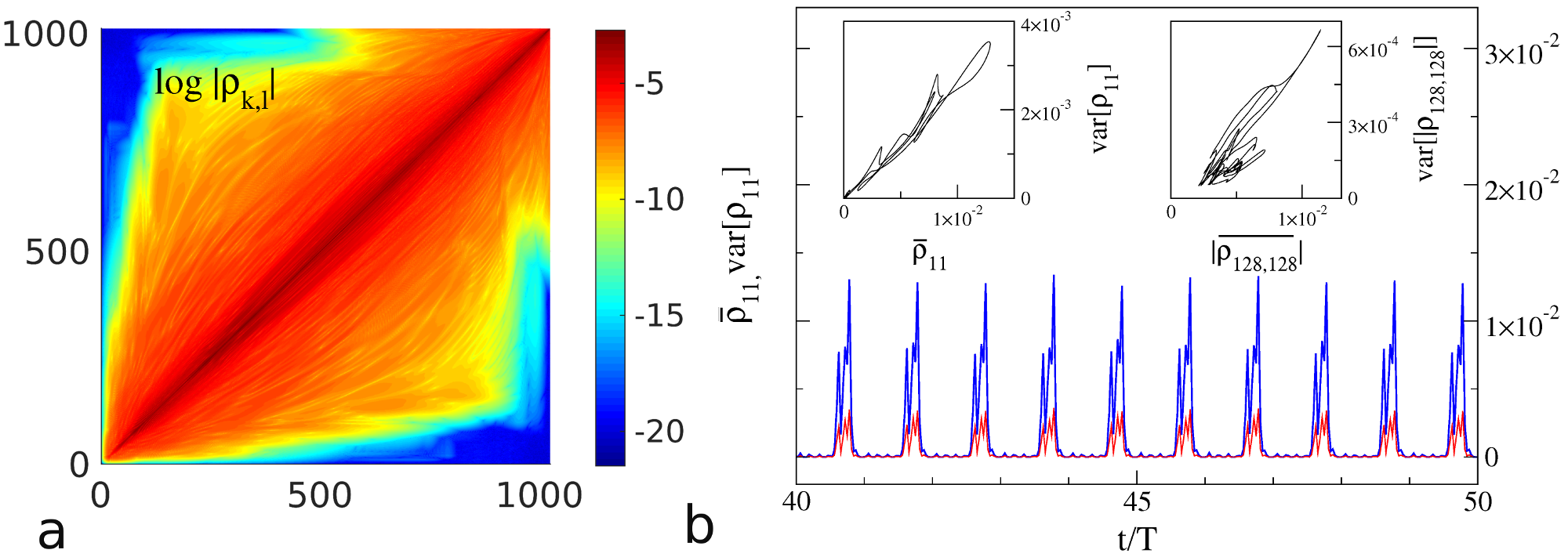}
\end{tabular}
\caption{(a) Stucture of the sampled stroboscopic density matrix
$\varrho^\mathrm{att}(0)$ ($N = 1024$) and (b) time evolution of the mean
$\bar{\varrho}_{11}(t)$ (thick blue line) and variance $\var[\varrho_{11}(t)]$
(thin red line) for $t \in[40T, 50T]$. The system size is $N=256$ and the sampling was
performed over $10^5$ independent trajectories initiated at the state
$\ket{\psi^{\mathrm{init}}} = \ket{1}$ and then propagated to the time
$t_{\mathrm{p}} = 50T$. The inset depicts the limit-cycle evolution of the means and
variances for two diagonal elements, $\varrho_{1,1}$ and $\varrho_{128,128}$,
during one period of modulations, $t \in [1000T, 1001T]$. Curves for later
periods are indistinguishable from the presented ones. The parameters are the
same as in Fig.~\ref{Fig:Husimi}.}
\label{Fig:3}
\end{figure*}

The efficient utilization of a supercomputer requires a reasonable scaling on
the distributed memory. In this regard, quantum trajectories possess an ideal
parallelization potential. The method realizes the general Monte Carlo paradigm
with independent simulations and without substantial load imbalance. The
transfer of the resulting data is the only data interchange between nodes. We
ran numerical simulations utilizing up to $32$ nodes of the supercomputer and
found that the implementation scales almost linearly with the number of nodes.
Next, we consider the performance and the scaling efficiency of the
implementation on $16$ CPU cores with shared memory. To start, the number of
MPI processes and OpenMP threads have to be chosen. We tried several different
configurations; namely, $1$ process $\times$ $16$ threads, $2$ processes
$\times$ $8$ threads, $4$ processes $\times$ $4$ threads, $8$ processes
$\times$ $2$ threads, and $16$ processes $\times$ $1$ thread. We did not find
a substantial difference in performance and chose the option $1~$ MPI process with
$16$ OpenMP threads mode for illustration.
It is known that setting a relevant affinity mask to pin threads to
CPU cores usually affects performance and scalability. In this regard, we used
the following settings:
\texttt{KMP\_AFFINITY = granulatiny = fine, scatter}. For all performance
measurements in this section we considered the model setup,
Eqs.~(\ref{eq:Hamiltonian}, \ref{eq:jump}), with 63 bosons (i.e., with dimension
$N = 2^6 = 64$) and $640$ trajectories. The results of our computational
experiments are summarized in Table \ref{table1}. Upon inspection this shows that our
implementation allows $87\%$ scaling efficiency on $16$ CPU cores with shared
memory.

Then, we ran the Intel VTune Amplifier XE profiler to find main time-consuming
parts of our implementation. As a result we found that the high-performance
implementation of the dense matrix-vector multiplication with $zgemv$ takes
more than $99\%$ of the total computation time. This in turn means that there
is no potential for further optimization of the code.

Finally, we estimate the computation time to propagate a single trajectory on
a single-core as a function of system size $N$, see Table \ref{table2}. For the model
specified by Eqs.~(\ref{eq:Hamiltonian}, \ref{eq:jump}) this time scales as
$N^3$; this is due to the multiplication of the quadratic scaling of a dense
matrix-vector multiplication and a linear scaling of the jump frequency. The
latter scaling is, however, model specific and may differ for other
models. Thus,  the overall computation time may vary substantially
with the type of Hamiltonian or/and dissipators under study. On top, the values
we present in Table \ref{table1} depend on the values of the coupling constant
$\gamma_0$ and the period of modulations $T$. This is so because these
parameters control the occurrence for the jumps. Therefore, these specific
estimates should not be identified as invariant quantifiers.

\begin{table}[b!!]
\caption{Single-core computation time to propagate a trajectory over one period $T$ as a function of $N$.
The parameters are $J=1$, $\mu_0 = 1.5$, $\mu_1 = 1$, $U=3$, $\gamma_0 = 0.1$, and $S=20$.}
\label{table2}
\vspace{6pt}
\small
\centering
\begin{tabular}{c c c c c c c c c}
\hline
Number of states, & Time of computations, \\
         $N$      & in seconds \\
\hline
\ 64       \ & 0.37 \\
\ 128      \ & 2.3 \\
\ 256      \ & 16 \\
\ 512      \ & 153 \\
\ 1024     \ & 1153 \\
\ 2048     \ & 8642 \\
\ 4096$^*$ \ & 64785 \\
\hline
$^*$ extrapolation
\end{tabular}
\end{table}

\section{Applications}

We now report the results of our simulations obtained for the model given by  Eqs.~(\ref{eq:Hamiltonian} - \ref{eq:jump}). We start with the performance
of the algorithm, Table~\ref{table1}. The idea of the algorithm mimics a float:
The algorithm constantly attempts to `float to the surface', i.e., to increase
the time step of integration towards its maximal value $\delta t_0$ while
every next jump pulls it downwards to $\delta t_S$, see Fig.~\ref{Fig:Husimi}. The average time
between two consequent jumps is the mean of the local maxima in the depicted
saw-like time sequence of $\delta t$. There is no problem in overestimating
$\delta t_0$, simply because the time step will rarely reach its maximum. The
shortest time step, $\delta t_S$, or, equivalently, the depth $S$, is tuned to
the values needed to reach the desired accuracy.

Next we turn to the averages $\bar{\varrho}^\mathrm{att}_{kl}(t)$ over realizations and the corresponding
statistical variances $\var[\varrho^\mathrm{att}_{kl}(t)]$ of the matrix
elements. Both quantifiers
converge to ``limit cycles'' if the propagation time $t_p=nT+\tau$, $n\in\mathbb{Z}^+$, $\tau\in[0,T)$, is much larger than all relaxation times.  This
means that for $n\gg1$ the density matrix converges to a time-periodic quantum attractor, i.e., $\bar{\varrho}^\mathrm{att}(t+T) = \bar{\varrho}^\mathrm{att}(t)$ [see Fig.~\ref{Fig:3}(a)]
and the variances also become time-periodic functions, $\var[\varrho^\mathrm{att}_{kl}(t+T)] =
\var[\varrho^\mathrm{att}_{kl}(t)]$ [see Fig.~\ref{Fig:3}(b)]. The crumpled caustic-like shapes of the limit cycles is a result of the projection on a plane of a 
global limit-cycle living in $N^4$-dimensional space: the limit-cycle are not topological products of $N^2$ two-dimensional limit cycles;
elements of the asymptotic density matrix do not evolve independently, they do interact so that their means and variances are correlated.

 In the asymptotic regime, the sampling can be performed
stroboscopically, i.e., after every period $T$. In our simulations we used  $t_{\mathrm{p}} = 1000T$ as the
transient time and then performed the stroboscopic sampling of $\varrho^\mathrm{att}(\tau=0)$. The attractor density 
matrix at any other instant of time $\tau \in [0,T]$ can be sampled by shifting the starting
time of the sampling, $t_{\mathrm{p}} \rightarrow t_{\mathrm{p}}+\tau$, or also
by performing an extra-sampling at all needed intermediate points.

\begin{figure}[b]
\includegraphics[width=0.48\textwidth]{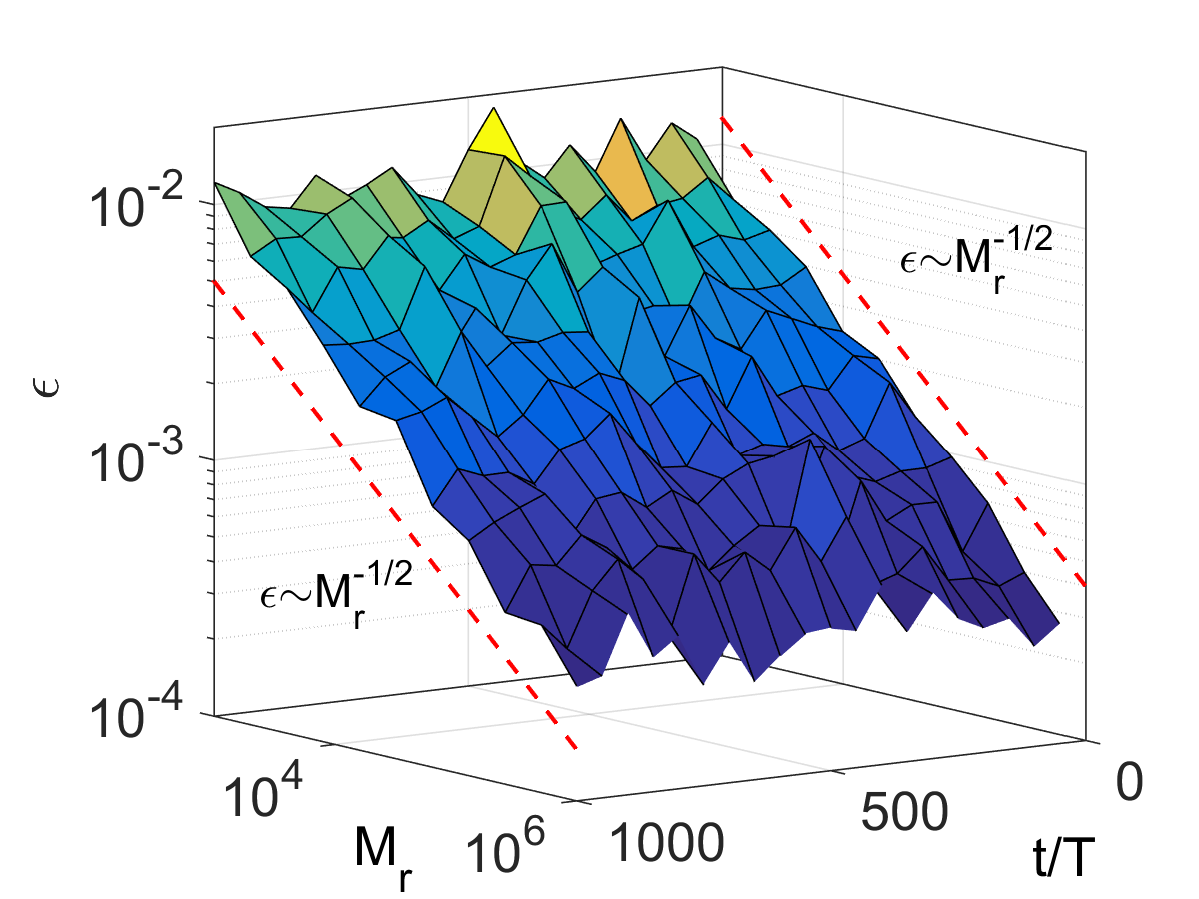}
\caption{Spectral norm of the difference between the density matrix 
stroboscopically sampled with quantum trajectory algorithm  and numerically 
exact asymptotic solution, $\epsilon=\left\|\bar{\varrho}^\mathrm{att}(mT)-\varrho^{ex}(\tau=0)\right\|$,
for the dimer model, Eqs.~(\ref{eq:Hamiltonian}, \ref{eq:jump}). Here $N = 100$, $J=1$, 
$\mu_0 = 1.5$, $\mu_1 = 1$,  $U=3$, $\gamma_0 = 0.1$.}
\label{Fig:3a}
\end{figure}

For relatively small system sizes, $N\simeq100$, we can obtain a 
numerically exact asymptotic solution, calculated as the kernel of  the 
Floquet map minus identity, $(\mathcal{P}_T-\mathbb{1})\varrho^{ex}(0)=0$. It 
allows us to quantify  convergence of the sampled density matrix -- with the icnrease of  the number of sampled 
trajectories, $M_r$, -- to the asymptotic state. The error is defined as the spectral 
norm of the  difference  matrix, $\epsilon=\left\|\bar{\varrho}^\mathrm{att}(mT)-\varrho^*(\tau=0)\right\|$ \cite{norm}. 
We find that, for the chosen set of parameters, the sampled solution converges to an attractor already  after $t_p >50T$, such that the observed error remains essentially time-independent. 
The resulting plot demonstrates that the sampling error scales 
as $1/\sqrt{M_r}$ (as expected for an independent Monte Carlo sampling) with no signatures of saturation; see Fig.\ref{Fig:3a}.

\begin{figure}[t]
\includegraphics[width=0.48\textwidth]{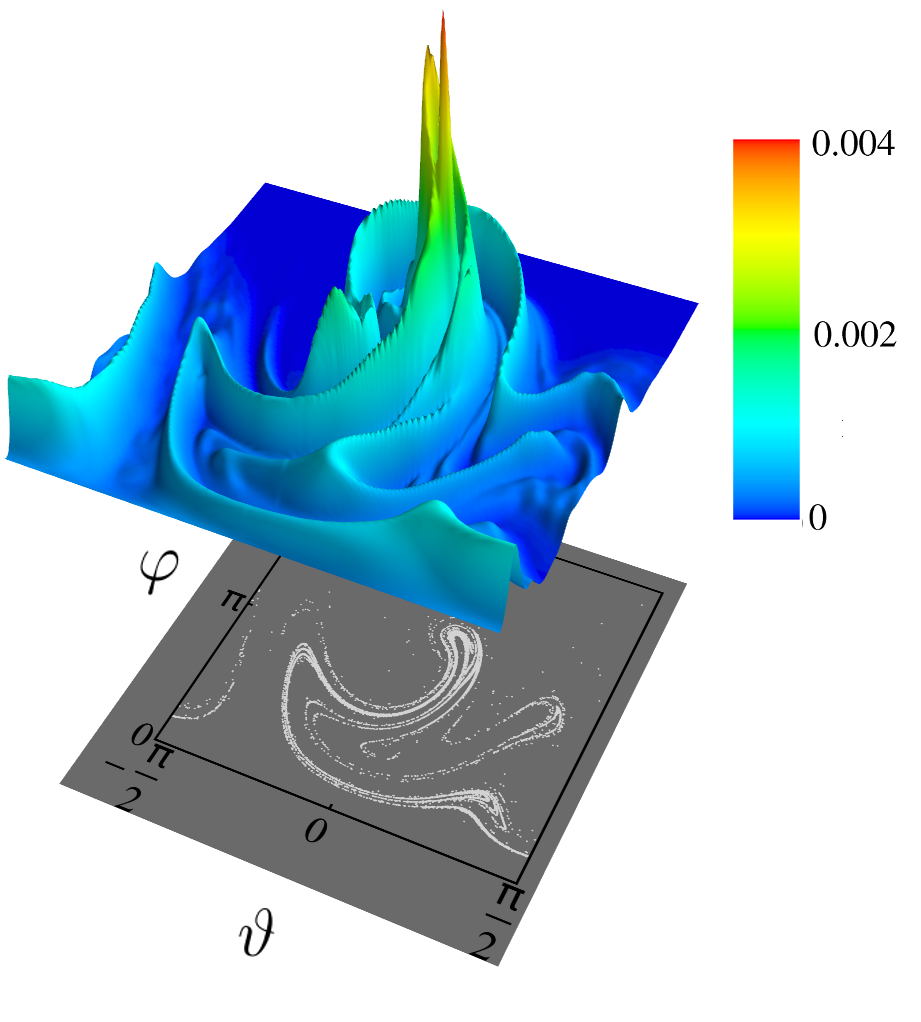}
\caption{Attractors of the dimer model, Eqs.~(\ref{eq:Hamiltonian},
\ref{eq:jump}). Husimi distribution of the attractor density matrix
$\varrho^\mathrm{att}(0)$ of the dimer with $N -1 = 1023$ bosons (top) and
classical attractor of the corresponding mean-field systems,
Eq.~(\ref{eq:mean}) (bottom panel). The density matrix was sampled with
$10^5$ stroboscopic realizations. The parameters are $J=1$, $\mu_0 = 1.5$,
$\mu_1 = 1$, $U=3$, $\gamma_0 = 0.1$.}
\label{Fig:4}
\end{figure}

With $4000$ samples per trajectory (that amounts to an additional propagation
for the time $4000T$) it became possible to collect $M_{\mathrm{r}} = 10^5$
samples for the model system of the dimension $N=1024$ (i.e., $N-1 = 1023$
indistinguishable bosons) by running the program on $32$ cores during three
days. 
The Husimi distribution of the sampled
density matrix is depicted in Fig.~\ref{Fig:4}. There is an intriguing similarity
between the distribution of the quantum attractor and the phase-space
structure of the classical attractor (i.e., its stroboscopic section, to be
more precise) produced by the mean-field equations. This allows us to conjecture that the attractor density
matrix was resolved with a good accuracy.
The $128$ cores allowed us to sample the
same number of realizations for the model with dimension $N=2048$ during
approximately one week \cite{foot5}.

\section{Conclusions}

The objective of this study was to estimate the numerical horizon of a
high-accuracy sampling of non-equilibrium dissipative  states of
periodically driven quantum systems  by using a high-precision realization of the quantum trajectory method. We demonstrated that, by implementing the algorithm 
on computer cluster with $\leq 128$ cores, it is possible to resolve time-periodic asymptotic density operator of
driven open quantum systems of several thousand of states  on a time scale of a few days. The
benefit of gaining access to the whole density matrix is the possibility to
extract more detailed information about the  non-equilibrium regimes which is encoded in the
matrix structure of the density operator, such as the purity and  many-body entanglement \cite{map}.

We would like to surmise on possible optimization  of the sampling
procedure. An immediate idea  is to use an optimal initial state $|\psi^{\mathrm{init}}\rangle$ in order to
reduce the transient time $t_p$. When it is about resolving the asymptotic density operator 
as a function of the value of a parameter changed within some range, the last moment wave vectors for the current parameter value 
can be used as the initial states to sample operator for the next parameter value. Next,
the performance of the algorithm can be substantially increased by 
grouping trajectories into matrices and substituting a
set of matrix-vector multiplications with a single matrix-matrix multiplication. 
Our tests have  shown that even in the presence of intrinsic asynchrony between different trajectories, 
this modification leads to a more than tenfold acceleration of the sampling process \cite{fold}.

Research areas where 'quantum attractors' are of potential interest have been
already mentioned in the introduction. Here, we like to recall them.
First, this is many-body localization \cite{alt} where the action of temporal
modulations \cite{abanin,lazarides} and dissipation \cite{fish,les} so far have
been considered separately. A combined action of both factors in MBL systems of non-regular topology presents an
intriguing challenge. The next issue is the survival of Floquet topological insulators \cite{top} in
the presence of dissipation or creation of new types of insulating Floquet states with
synthetic dissipators,  are
timely objectives of interest for practical applications.
Finally, our numerically exact realization of quantum trajectory method can be used 
to explore - in a very accurate way - the thermodynamics of quantum jump trajectories \cite{jumpL}
in complex periodically-modulated  open quantum systems and search for non-equilibrium analogs of dissipative phase transitions \cite{PT}.

\section{Acknowledgments}\label{acknowledgment}

The authors acknowledge support of the Russian Science Foundation grant No.
15-12-20029. 
The authors also thank I. Vakulchik for the help with preparing the figures.


\begin{thebibliography}{1}
\bibitem{foot1} 
Even if the Hamiltonians $H_1$, $H_2$ and the Lindblad operators
$\mathcal{L}_1$,$\mathcal{L}_2$ are sparse; thus the Floquet map $\mathcal{P}(T) =
e^{\mathcal{L}_2(T-\tau)}e^{\mathcal{L}_1\tau}$ yields a dense matrix. Therefore,
the numerical evaluation cannot benefit from sparse matrix methods.
\bibitem{foot2} 
There is a sophisticated way to
propagate Eq.~(\ref{lind}) by using the time evolving block
decimation (TEBD) technique \cite{cirac1,vidal}. 
The corresponding numerical effort scales as $\log
N$. However, this algorithm can only be used for systems of lattice topology, i.e., systems
that can be partitioned into $\sim \log(N)$ `pieces' coupled by next-neighbor
interactions, both unitary and dissipative. In this case it yields the correct
answer when the asymptotic state is characterized by a short-range entanglement.
\bibitem{foot3} A generalization of the TEBD technique to non - Hermitian operators 
was used for
the propagation step in Refs.~\cite{TEBD1,TEBD2}. Similar to its Hermitian
predecessor, that method can only be implemented for lattice systems.
Also, it guarantees correct asymptotic results only when there no distant entanglement
build-ups in the course of the system evolution.
\bibitem{foot4} 
The components of the initial state vector 
are taken independent and identically uniformly distributed in $[-1/2,1/2]$, and then the vector is normalized to unity.
\bibitem{foot5} Calculations of the Husimi distribution for the
model (\ref{eq:Hamiltonian}-\ref{eq:jump}) involve summation over series of
binomial coefficients of the order $N$. It was not possible for us to go
beyond $N=2^{10}=1024$ when calculating the corresponding Husimi
distributions.
\end{thebibliography}
\end{document}